# Electronic states of *trans*-polyacetylene, poly(*p*-phenylene vinylene) and *sp*-hybridised carbon species in amorphous hydrogenated carbon probed by resonant Raman scattering

M. Rybachuk [1, 2 *] and J.M. Bell [2]

[1] *Federal Institute for Materials Research and Testing (BAM), Division VI.4 Surface Technology, Unter den Eichen 87, 12205 Berlin, Germany*

[2] *Faculty of Built Environment and Engineering, Queensland University of Technology, 2 George St, Brisbane, Qld 4001, Australia*

**Abstract**

Inclusions of *sp*-hybridised, *trans*-polyacetylene [*trans*-(CH)$_x$] and poly(*p*-phenylene vinylene) (PPV) chains are revealed using resonant Raman scattering (RRS) investigation of amorphous hydrogenated carbon (*a*-C:H) films in the near IR – UV range. The RRS spectra of *trans*-(CH)$_x$ core $A_g$ modes and the PPV *CC-H* phenylene mode are found to transform and disperse as the laser excitation energy $\hbar\omega_L$ is increased from near IR through visible to UV, whereas *sp*-bonded inclusions only become evident in UV. This is attributed to $\hbar\omega_L$ probing of *trans*-(CH)$_x$ chain inhomogeneity and the distribution of chains with varying conjugation length; for PPV to the resonant probing of phelynene ring disorder; and for *sp* segments, to $\hbar\omega_L$ probing of a local band gap of end-terminated polyynes. The IR spectra analysis confirmed the presence of *sp*, *trans*-(CH)$_x$ and PPV inclusions. The obtained RRS results for *a*-C:H denote differentiation between the core $A_g$ *trans*-(CH)$_x$ modes and the PPV phenylene mode. Furthermore, it was found that at various laser excitation energies the changes in Raman spectra features for *trans*-(CH)$_x$ segments included in an amorphous carbon matrix are the same as in bulk *trans*-polyacetylene. The latter finding can be used to facilitate identification of *trans*-(CH)$_x$ in the spectra of complex carbonaceous materials.

[*] Corresponding author. Fax: +61 7 3138 4135. E-mail address: m.rybachuk@qut.edu.au
(M. Rybachuk)

# 1. Introduction

Amorphous carbon (*a*-C) and diamond-like carbon (DLC) solids are characterised by a large variety of types and properties that stem from combinations of principally two hybridised forms of carbon $sp^2$ and $sp^3$ and, for carbon materials formed in presence of hydrogen, as for *a*-C:H, the resultant properties are also controlled by the hydrogen content. Isotropic materials like DLC or *a*-C can, in principal, contain inclusions of a basic polymer, the *trans* isomer of polyacetylene [*trans*-(CH)$_x$] according to simulations by Bernasconi *et al.* [1]. This introduced the idea that *C–C* bonds in bulk *trans*-(CH)$_x$ undergo a gradual saturation via chain interlinking at high pressure, transforming into an *a*-C:H solid, and on the other hand, earlier experiments by Arbuckle *et al.* [2] showed that $sp^3$ clustering occurs if defect concentrations in *trans*-(CH)$_x$ reach sufficiently high level. These findings can be related to the energetic mechanism of $sp^2$ and $sp^3$ bonding formation in a hydrogenated DLC [3]. The presence of *trans*-(CH)$_x$ in a carbonaceous solid was reported by López-Ríos *et al.* [4] for CVD synthesised diamond, and Dischler *et al.* [5] and Piazza *et al.* [6] identified *trans*-(CH)$_x$ inclusions in low temperature synthesised *a*-C:H. Assignment of a Raman peak at *ca.* 1140 cm$^{-1}$ to $\omega_1$ *C–C* in plane bending mode and a peak at *ca.* 1490 cm$^{-1}$ to $\omega_3$ *C=C* stretching mode to those of *trans*-(CH)$_x$ was, at first, uncertain, since solution synthesised *trans*-(CH)$_x$ is known to be unstable at the elevated temperatures used in ordinary DLC deposition [4, 7, 8]. Some authors inferred that very short (less than 20 *C=C* units) temperature stable *trans*-(CH)$_x$ segments are formed between the diamond grains during deposition [4, 9, 10]. Isotopic substitution experiments by Kuzmany *et al.* [11] and Michaelson *et al.* [12] confirmed the assignment of the $\omega_1$ and $\omega_3$ modes to *trans*-(CH)$_x$. Recently Teii *et al*. [13] made an effort to correlate the interaction between the hydrogen-rich plasma and the amount of *trans*-(CH)$_x$ in nanocrystalline diamond thin-films. Conclusive results, however, were not obtained since at present, the means to quantitatively identify the amount and/or the ordering of *trans*-(CH)$_x$ inclusions in a given *a*-C or DLC solid are not sufficiently defined.

The purpose of this work is to present the resonant Raman scattering (RRS) investigation of basic *a*-C:H films in the near-infrared (NIR) to ultraviolet (UV) range, and to demonstrate that these films host *trans*-(CH)$_x$ inclusions (chains) characterised by intrinsic ordering and variable conjugation length; and to show that films also contain *sp*-hybridised carbon species and inclusions of poly(*p*-phenylene vinylene) (PPV) [14]. The *sp*-bonded species considered are short hydrogen-terminated polyyne chains. Here we obtain experimental and theoretical results that demonstrate differentiation between the Raman modes of *trans*-(CH)$_x$ (core $A_g$ modes) and the PPV phenylene mode in *a*-C:H. We illustrate that at various laser excitation energies ($\hbar\omega_L$), the changes in Raman spectra features for *trans*-(CH)$_x$ segments included in an amorphous carbon matrix of an *a*-C:H are the same as in bulk *trans*-polyacetylene. Figure 1 shows an example of the RRS for bulk *trans*-polyacetylene for different excitation laser wavelengths ($\omega_L$). This figure shows as the $\hbar\omega_L$ is increased from deep-red to blue excitation, the RRS bands change gradually from narrow, slightly asymmetric lines into more complex two-peak bands, each consisting of an un-shifted primary peak and an upward shifted satellite portion which becomes the prominent feature of the band at blue excitation [15]. In *a*-C:H too, the $\omega_1$ and $\omega_3$ *trans*-(CH)$_x$ modes become transformed and change positions, (disperse [9]), shapes and intensities with changing $\hbar\omega_L$, and these transformations strongly depend on the inherent degree of inhomogeneity of *trans*-(CH)$_x$ chains.

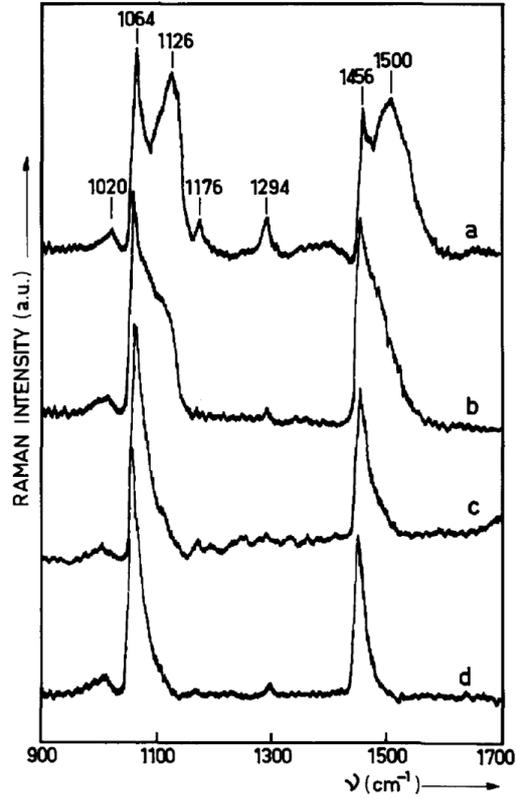

Fig. 1 - Resonant Raman spectra of bulk *trans*-(CH)$_x$ at 78 K taken for different laser excitation wavelengths. a) $\omega_L$ = 457.9 nm; b) $\omega_L$ = 514.5 nm; c) $\omega_L$ = 600 nm; d) $\omega_L$ = 676.4 nm, adapted from Ref. 15.

The distribution of *trans*-(CH)$_x$ segments with varying degrees of inhomogeneity (conjugation chain length, bond disorder) was computed employing the bi-modal chain distribution model proposed by Brivio and Mulazzi *et al.* [8, 15, 16] and the amplitude mode theory proposed by Ehrenfreund *et al.* [7]. We aim to elucidate a simple approach that is needed to facilitate extraction of *trans*-(CH)$_x$ contributions from core Raman spectra of *a*-C, DLC or any other carbonaceous materials.

The RRS has been successfully used to study inhomogeneity and disorder in amorphous carbon systems consisting of arbitrary combinations of *sp*, $sp^2$ and $sp^3$ hybridised states [3, 7, 9, 17-19]. In this work the excitation energies over a wide range of energies from 1.58 eV to 5.08 eV were used, ensuring that the vibrational densities of states (VDOS) of the great majority of $sp^3$, $sp^2$ and *sp* carbon mixtures are measured.

In an environment where energetic hydrogen ions are present, the probability for carbon atoms to enter into *sp*-type arrangement as either polyyne $(-C\equiv C-)_n$, a semiconductor, or polycumulene $(=C=C=)_n$, a semi-metal [20], is exceedingly low, since both species are highly unstable to hydrogen exposure [21] and temperature sensitive [5, 22, 23]. Identification of *sp*-hybridised inclusions in the hydrogenated $sp^2$–$sp^3$ aggregates are therefore, highly notable, since the *sp* self-organisation mechanism, even at present, remains largely unresolved [24-28]. We focused on detection of *sp*-hybridised segments using the RRS process, and critically consider recent findings by D'Urso *et al.* [19]. In addition, infra-red (IR) absorption spectra analysis was used to identify *sp* and *trans*-$(CH)_x$ species.

## 2. Experimental

Deposition of *a*-C:H films were performed on *Si* <111> wafers using a Helmholtz-type inductively coupled plasma (ICP) reactor operated on $CH_4$/*Ar* mixture at temperatures of 380 – 400 K [29, 30]. The pressure was $6 \times 10^{-2}$ Pa and the substrate was DC negatively biased at -250 to -300 V. The use of substrate bias this range was found to have adverse effects to the formation of *sp* or *trans*-$(CH)_x$ segments. The formation of *trans*-(CH)$_x$ and *sp* bonded segments appears to be facilitated by the amount of atomic hydrogen in hydrocarbon plasma, and the ratio of 55% *CH$_4$* to 45% *Ar* was found to be most favourable in this work. Deposition was performed at extremely low rate of ~30 nm/hour in a high density plasma, with the aim of obtaining high ordering of *sp$^2$* phase, and allowing for a higher concentration of free radicals and a higher degree of gas phase reaction taking place.

The fabricated films were found to be of low intrinsic compressive stress ≤ 1 GPa as determined using Stoney's equation [31] from the substrate curvature with hardness of approximately 20 GPa and, a friction coefficient of 0.07 at 70% humidity measured by a nano-mechanical testing (UMIS). Electrical resistivity was in the range of $10^8$ - $10^9$ Ω cm$^{-1}$ as measured by using a four-probe testing method. Films were ~140 nm thick with a maximum refractive index of 2.2 in the UV – blue region, as measured by IR – UV spectroscopic ellipsometry (J. A. Woollam Co.). The hydrogen content in the films was determined from the analysis of IR absorption spectra as used by Liu *et al*. [32] (normal mode vibrational frequency calculations) and from the analysis of UV Raman spectra ($\omega_L$ of 244 nm) as proposed by Casiraghi *et al*. [33] where the full width at half maximum (FWHM) of the *G* peak, *G* peak position and the dispersion of this peak at respective $\hbar\omega_L$ were used. This gave a hydrogen content of approximately 27 (± 2.5) at%.

The IR spectra in the range 3400 – 2600 cm$^{-1}$ range were obtained using Nicolet Nexus Fourier transform infra-red (FT – IR) spectrometer operated in transmission mode with subtraction of *Si* substrate background. For IR measurements, the same group (thickness, lattice orientation, and surface finish grade and backside surface roughness) of uncoated *Si* substrates was used, as for the film deposition experiments.

Standard Gaussian peak functions were used to fit the constituent bands in the selected spectral range after linear background subtraction.

X-ray photoelectron spectroscopy (XPS), using Kratos AXIS Ultra photoelectron spectroscope with a monochromated Al $K_\alpha$ 1486.6 eV X-ray source, was used *ex situ* to obtain $C_{1s}$ spectra. The chamber vacuum level was maintained below $2.5 \times 10^{-9}$ Pa and the spectrometer was calibrated by peak referencing of Au $4f_{7/2}$ (binding energy = 84.0 eV) with respect to the Fermi level. XPS measurements were collected centred at 284.0 eV at pass energy of 40 eV, with a resolution of 0.05 eV and dwell time of 250 msec; a total of 3 collection sweeps were used. Charge neutraliser was off and surface charging was not observed during the measurements. Information about the relative abundance of carbon hybridised fractions in the examined *a*-C:H materials was obtained by decomposition of $C_{1s}$ core electron binding energy spectra onto three constituent peaks: *sp*, *sp$^2$* and *sp$^3$*. After the subtraction of Shirley background [34], Pearson VII line-functions corresponding to these peaks were fitted into the main $C_{1s}$ peak employing the constrained fitting procedure, where an *sp* constituent was fitted with an assigned binding energy (position) [35, 36] and, *sp$^2$* and *sp$^3$* constituents were fitted restricted to their respective energy separation gap [37] then, the value ratio of *sp*/*sp$^2$*/*sp$^3$* was obtained based on integrated peak areas for the three hybridised line-functions. The presence of *sp*-hybridized species revealed by means of the XPS $C_{1s}$ analysis was verified by analysing 244 nm Raman [19, 38, 39] and IR [5] results, and the *sp$^3$* content by using 244 nm Raman results [17, 33].

Unpolarised Raman spectra across the excitation energy range 5.08 eV to 1.28 eV were obtained *ex situ* at 293 K using 244, 532, 633, and 785 nm Renishaw instruments and 325 and 442 nm Kimmon Raman instruments. All excitation wavelengths excluding 785 nm were pulsed; the 785 nm was a continuous wavelength laser source. The frequency-doubled *Ar* ion laser was used for 244 nm, *He*/*Cd* for 325 and 442 nm, the frequency-doubled YAG laser was used for 532 nm, *He*/*Ne* gas laser was used for 633 nm, and a diode laser source was used for 785 nm excitations. All measurements were taken in dynamic mode with a specimen moved linearly at speeds of up to 30 m/s and laser power was kept at or below 1 mW for all wavelengths minimizing the thermal damage. The acquisition time was varied between 10 s to 120 s and the spectral resolution was 1 $cm^{-1}$.

There were two main options for fitting of the Raman spectra as noted by Casiraghi *et al*. [33]: an all Gaussians fit to Raman constituent bands, or a fit with a Breit–Wigner–Fano (BWF) line shape for the *G* peak and a Lorentzian for the *D* peak. The *a*-C:H samples selected for this study did not display significant photoluminescence (PL) background, nonetheless, we find that the use of the BWF line is not the most appropriate since the BWF *Q* coupling coefficient is influenced by the PL background, and the BWF lineshape tends to adjust its asymmetry reproducing a part of the PL slope [33, 40]. This does not lead to reproducible fitting of the Raman spectra. Alternatively, fully symmetric Gaussian line-shapes provide better, reproducible fit in the presence of a PL background. In the Raman spectra presented, the linear PL background was subtracted and all constituent peaks were fitted with Gaussian line-shapes using a nonlinear least squares fitting procedure [41].

**3. Results and discussion**

*3.1. Identification of π- conjugated polymeric inclusions in a-C:H*

Figure 2 shows the RRS spectra of an examined *a*-C:H film with PL background subtracted and fitted with Gaussian line-shapes to the constituent peaks.

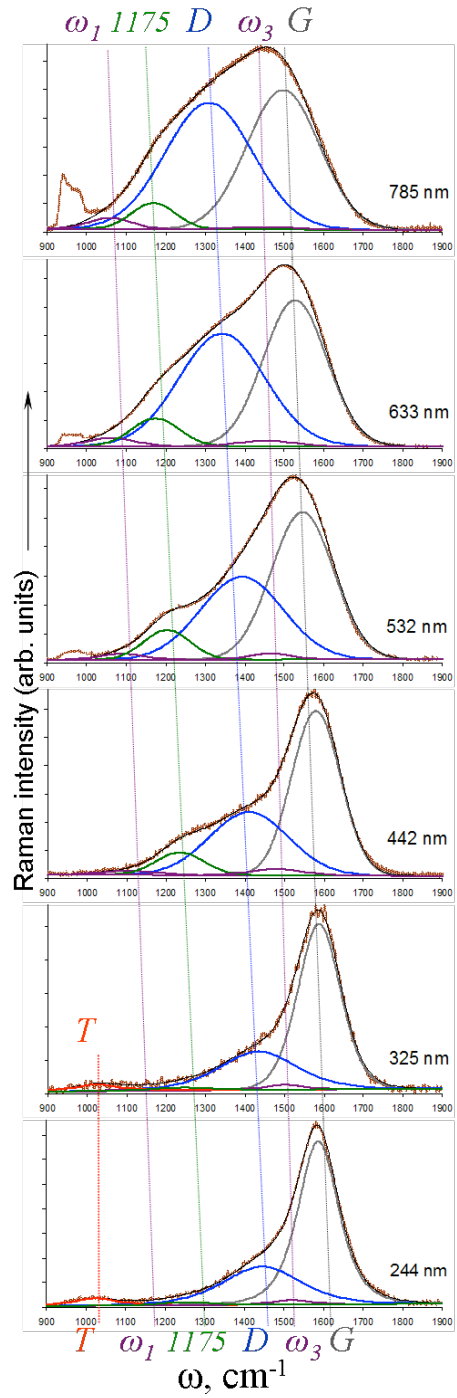

Fig. 2 - Resonant Raman spectra of *a*-C:H at 293 K showing contributions from *trans*-(CH)$_x$ ($\omega_1$ and $\omega_3$) modes, PPV (1175 cm$^{-1}$ mode), and DLC (*D*, *G*, and *T* modes). An asymmetric peak visible at NIR–visible (green) $\hbar\omega_L$ at 950 cm$^{-1}$ is the second order *Si*, from Ref. 14.

The fitted bands are common DLC *D* and *G* modes for NIR and visible and the *T* mode for UV $\hbar\omega_L$ excitations [17]; and the two $A_g$ zone center vibrational modes of *trans*-(CH)$_x$, the $\omega_1$ and $\omega_3$ [7, 18, 42]. The weak $\omega_2$ mode that usually appear at 1275 – 1295 cm$^{-1}$ range (a peak labelled '1294', Fig. 1) was not present, nor the peak corresponding to the $B_g$ mode of *trans*-(CH)$_x$ that is normally observed in 1000 – 1100 cm$^{-1}$ range (a peak labeled '1020', Fig. 1), however their contributions could be hidden by the tails of the fitted *D* and the $\omega_1$ bands. Fundamentally, the absorption for bulk *trans*-(CH)$_x$ occurs within 1.5 - 1.7 eV range and corresponds to the zone centre $A_g$ Raman modes at opening frequencies of 1060, 1280 and 1450 cm$^{-1}$ [18, 42]. That is at N-IR $\hbar\omega_L$. As the Raman excitation energy increases, and therefore moves away from the band gap resonance, the Raman sidebands exhibit radical lineshape changes as illustrated in Fig. 1 [15, 18, 43, 44]. Shoulders appear at the high frequency side of the primary $\omega_1$ and $\omega_3$ modes that eventually extend into secondary peaks at excitation energies well above the band gap at $\hbar\omega_L$ = 2.71 eV [18]. The RRS spectra disperses [9] and the resultant *trans*-(CH)$_x$ peaks change intensities, *I* and widths *Γ*, and the overall spectrum is in addition affected by light polarisation [45, 46]. The RRS of *trans*-(CH)$_x$ secondary peaks, such as the peaks appearing at $\omega_L$ of 457.9 nm (noted as '1126' and '1500', Fig. 1) in bulk samples become more pronounced at higher excitation energy. However, the complexity of separating *trans*-(CH)$_x$ from the host DLC modes leads us to analyse a single symmetric band distribution [43]. This approach was proved by Ehrenfreund *et al.* [7] to be sufficient to account for a double peak Raman structure.

Together with common DLC and *trans*-(CH)$_x$ modes we find a peak positioned at 1175 cm$^{-1}$ when probed by 785 nm $\omega_L$ which we assign to a *CC–H* bending mode of the ring in neutral poly(*p*-phenylene vinylene) (PPV) [47, 48]. The origin of this mode could, in fact, be due to defects in *sp*$^2$ aromatic rings since in single crystals, only phonons with the wave vector rule *k*=0 contribute to Raman scattering. Defects lead to relaxation of this selection rule and therefore provide means for phonons from outside the centre of the Brillouin zone to contribute to the Raman scattering. If this 1175 cm$^{-1}$ mode indeed belongs to PPV chains, the other PPV zone centre vibrational modes found at higher frequencies in 1200 – 1330 and 1540 – 1625 cm$^{-1}$ range will certainly be obscured by the host *D* and the *G* modes [49]. Owning to its large width, $\Gamma_{1175}$ the 1175 cm$^{-1}$ vibrational mode could be effectively a combination of vinylene and a *CC–H* ring bend modes since the zone mode frequency for vinylene is at approximately 1145 cm$^{-1}$ [47]. As the Raman excitation energy increases from NIR to UV range all peak positions shift to a higher vibrational frequency obeying phonon confinement rules [17], as shown in Figure 3(a) where peak dispersion, $\Delta\omega$ is denoted as the shift in a peak position relative to base position at NIR excitation ($\hbar\omega_L$=1.58 eV). Figure 3(b) summarizes changes in widths for all fitted peaks. The gradual decrease in *I(D)/I(G)*, the intensity ratio for the *D* and *G* peaks, from ~0.9 to 0.2, the pronounced reduction in $\Gamma_D$ and $\Gamma_G$, and the *G* peak saturating [17] at approximately 1590 cm$^{-1}$ for 244 nm excitation indicate that *a*-C:H films hosting *trans*-(CH)$_x$ and PPV inclusions consist of a highly ordered and symmetric *sp*$^2$ phase [17, 30].

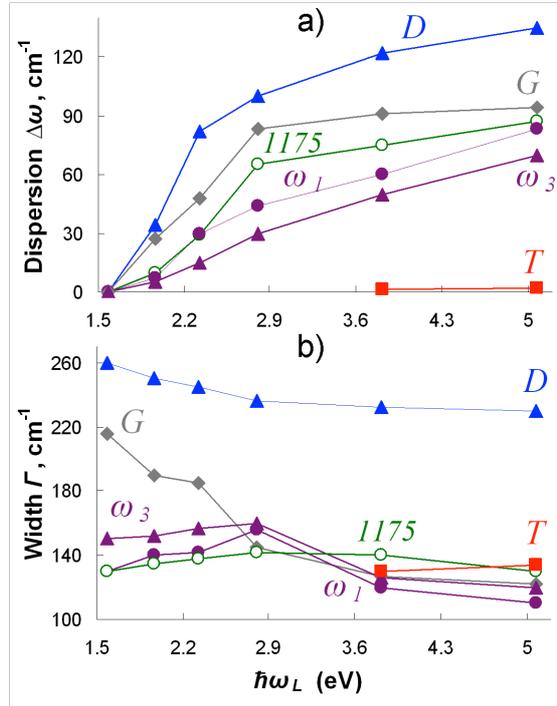

Fig. 3 a) - Peak dispersion, $\Delta\omega$ and b) - peak widths, $\Gamma$ for all constituent peaks as a function of the laser excitation energy $\hbar\omega_L$, from Ref. 14.

There is no *T* peak dispersion at higher excitation energies in agreement with earlier reports [17]. The band gap for PPV is in the range of 2.2 – 2.5 eV [48, 50] and therefore it is selectively probed by excitation energy corresponding to green light ($\omega_L$ of 532 nm). Figure 4 illustrates relative changes of the fitted $\omega_1$, $\omega_3$ and *1175* cm$^{-1}$ bands in the spectra of *a*-C:H as a function of $\hbar\omega_L$.

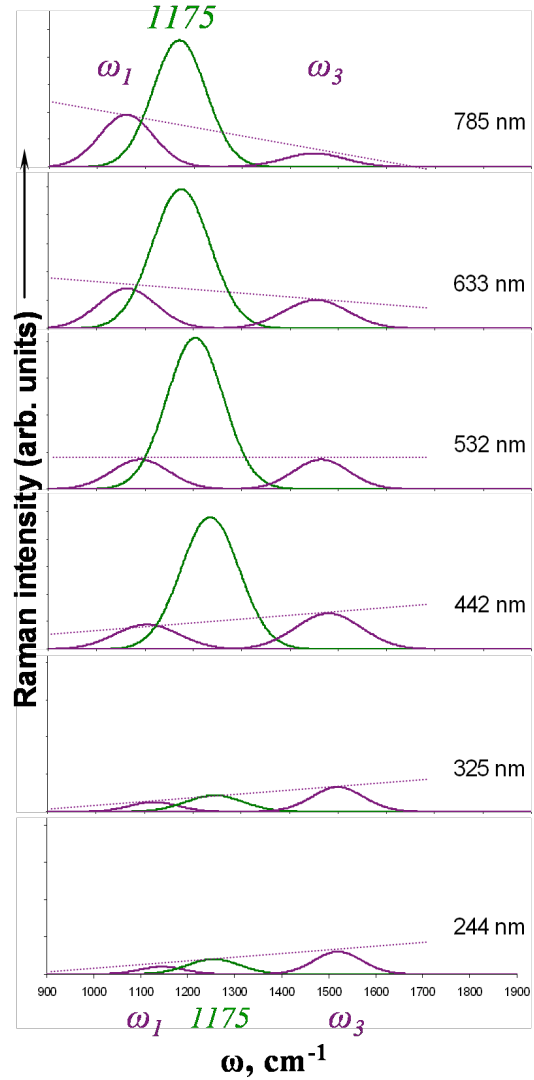

Fig. 4 - Changes in the fitted $\omega_1$, $\omega_3$ and *1175* cm$^{-1}$ bands (intensity magnified by a factor of 5) relative to the laser excitation energy $\hbar\omega_L$ in the spectra of *a*-C:H. Light dotted line over $\omega_1$ and $\omega_3$ bands denotes the $I(\omega_3)/I(\omega_1)$ trend.

Figure 5(a) illustrates changes in the relative intensity of the 1175 cm$^{-1}$ peak, *I(1175)* at different excitation energies; *I(1175)* is calculated as intensity of the *1175* peak over the total intensity of all constituent peaks including the intensity of the *T* peak in the UV Raman spectra.

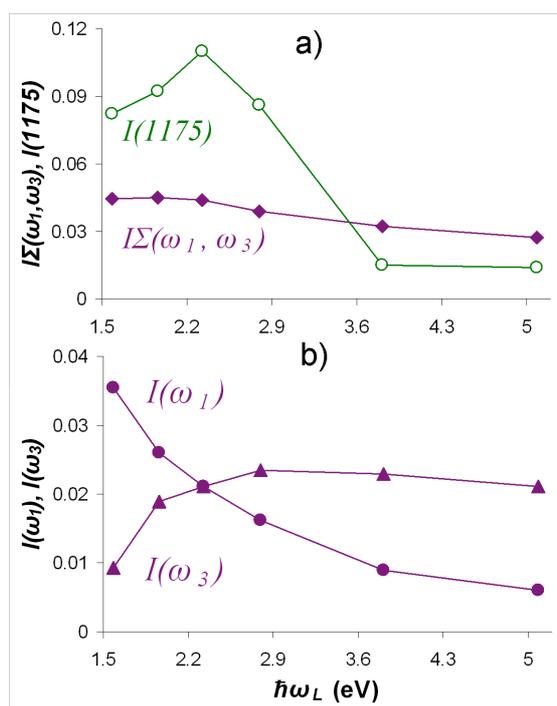

Fig. 5 a) - Evolution of relative intensities of the *1175* cm$^{-1}$ peak, $I(1175)$ and *trans*-(CH)$_x$ contributions, $I\Sigma(\omega_1, \omega_3)$ and b) relative intensities of $\omega_1$, $I(\omega_1)$ and $\omega_3$, $I(\omega_3)$ peaks as a function of the laser excitation energy $\hbar\omega_L$.

The $I(1175)$ reaches its highest intensity position at $\hbar\omega_L$=2.3 eV as revealed in Fig. 4, although Fig. 3(b) shows that changes in the peak width, $\Gamma_{1175}$ are minor at this band bap frequency. This PPV peak at 1175 cm$^{-1}$ is certainly of $sp^2$ origin since its contributions disappear in UV excitation.

The total relative intensity for *trans*-(CH)$_x$ contributions calculated as sum of relative intensities of $\omega_1$, $I(\omega_1)$ and $\omega_3$, $I(\omega_3)$ peaks and denoted as $I\Sigma(\omega_1, \omega_3)$ is shown in Fig. 5(a). The magnitude of $I\Sigma(\omega_1, \omega_3)$ gradually decreases from N-IR to UV $\hbar\omega_L$. Individual trends of $I(\omega_1)$ and $I(\omega_3)$ contributions are shown in Fig. 4 and Fig. 5(b). The *trans*-(CH)$_x$ features as $I(\omega_1)$ and $I(\omega_3)$ intensities (Fig. 4 and Fig. 5(b)) and peak widths, $\Gamma$ (Fig. 3(b)) for the $\omega_1$ and $\omega_3$ peaks exhibit strong transformations as the excitation energy increases from NIR excitation to blue excitation, but is most pronounced with excitation energy corresponding to green light; that is $I(\omega_1)$ and $I(\omega_3)$ intensities becoming inversely related with a fall in the total intensity, $I\Sigma(\omega_1, \omega_3)$ and $\Gamma_{\omega 1}$ and $\Gamma_{\omega 3}$ achieve a maximum in the blue–green excitation region. These are not related to tuning into the band gap frequency for *trans*-(CH)$_x$ that requires much less energy (in the NIR [18, 42]), but have been regarded as evidence of the presence of inhomogeneity (disorder) in *trans*-(CH)$_x$ chains.

The disorder is due to a distribution of the electronic energy gaps and their respective frequencies; these are selectively probed via the variation of $\hbar\omega_L$ of RRS process, and result in the shift and broadening of phonon bands [7]. Major attempts to describe the inhomogeneity of *trans*-(CH)$_x$ via a distribution of chains with varying length of $\pi$-electron conjugation [51] employed a particle-in-the-box [52] and Hückel-type calculations [53]. The bi-modal distribution model proposed by Brivio and Mulazzi *et al.* [8, 16] suggested a double peak distribution to arise from individual contributions of both long and short *trans*-(CH)$_x$ segments that show unequal resonant enhancement at a given excitation energy. The inhomogeneity of *trans*-(CH)$_x$ could also be described employing the distribution of the electron-phonon coupling constant $\lambda$, $p(\lambda)$ [7, 54] of the amplitude mode (AM) theory proposed by Ehrenfreund *et al.* [7]. When other parameters are fixed, $\lambda$ determines the Peierls relation for the energy gap $E_g(\lambda) = W\ exp(-1/(2\lambda))$ and laser frequencies, $\omega_L$, and where $W$ is the width of the $\pi$-band. The maximum for $p(\lambda)$ occurs at $\lambda = \lambda_0$, whereas resonance induced changes in peak position, $I$ and $\Gamma$ result from the condition $\hbar\omega_L = E_g(\lambda) > E_g(\lambda_0)$. We applied the AM model to study *trans*-(CH)$_x$ inclusions in *a*-C:H and the obtained results yielded $\lambda$ distribution range from ~0.17 for NIR to ~0.24 for UV, in good agreement with the AM model. The distribution of $\lambda$ rises from finite localisation lengths and bond length disorder, consequently AM theoretical calculations indicate that *trans*-(CH)$_x$ segments probed by higher $\hbar\omega_L$ are of shorter $\pi$- conjugation lengths and of higher bond disorder. The AM findings are complemented by calculations determining the length of the segments using Brivio and Mulazzi bi–modal distribution model which offers empirical relations for dependence of the conjugation length (long and short) on the frequency of the $\omega_1$ and $\omega_3$ modes. It was found that the approximate length for both single *C–C* and double *C=C* bonds in probed *trans*-(CH)$_x$ segments is no less than 120 bond lengths units at the estimation limit of the model; and to a minimum of approximately 8. Shorter chains are probed by higher excitation energies, as illustrated in Figure 6 [16] for the functional dependence of the $\pi$- electron gap, $\Omega$, eV and the relative optical absorption on the chain lengths of *trans*-(CH)$_x$ segments.

The average chain population is ~25 (± 5) bond length units owing to the uncertainties given by the Raman fitting and the bimodal distribution model [8]. All *trans*-(CH)$_x$ chains included in *a*-C:H are highly disordered as evidenced by wide $\omega_1$ and $\omega_3$ Raman peaks reaching their maximum in the blue-green range, shown in Fig. 3(b).

We have calculated the theoretical distribution for $I(\omega_3)/I(\omega_1)$ vs. $\hbar\omega_L$ independent of a given *trans*-(CH)$_x$ chain length using the AM formalism that was previously completed by Ehrenfreund *et al.* [7] for the visible laser excitation to include N-IR and UV $\hbar\omega_L$. Fitting of $\omega_1$ and $\omega_3$ spectral constituents delivered the inverse relationship between the $I(\omega_3)$ and $I(\omega_1)$ parameters, $I(\omega_3)/I(\omega_1)$ relative to the laser excitation energy as illustrated in Fig. 4 and Fig. 5(b). Figure 7 shows that our experimental results obtained for the varying $\hbar\omega_L$ are in good agreement with the theoretical distribution predicted by the AM model and with Ehrenfreund's experimental data. The observation of relative intensities of $\omega_1$ and $\omega_3$ bands, $I(\omega_1)$ and $I(\omega_3)$ (Fig. 4), and the theoretical distribution of $I(\omega_3)/I(\omega_1)$ ratio (Fig. 7) relative to $\hbar\omega_L$ evidences that $\omega_1$ and $\omega_3$ peak resonance responses to excitation energy of green laser are essentially equal. These are indicated by equal magnitudes of $I(\omega_1)$ and

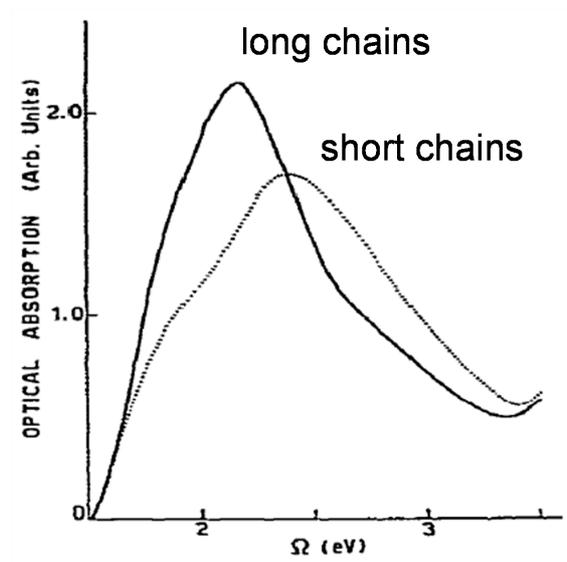

Fig. 6 - Calculated absorption spectra from long and short chains constituting *trans*-$(CH)_x$ samples, adopted from Ref. 16.

$I(\omega_3)$ shown in Fig. 4 and the $I(\omega_3)/I(\omega_1)$ ratio approaching 1.0 following the AM calculations graphically represented in Figure 7. This observation suggest that green Raman laser could become a wavelength of choice for natural identification of *trans*-$(CH)_x$ inclusions in carbonaceous materials fitting both $\omega_1$ and $\omega_3$ contributions at equivalent intensities.

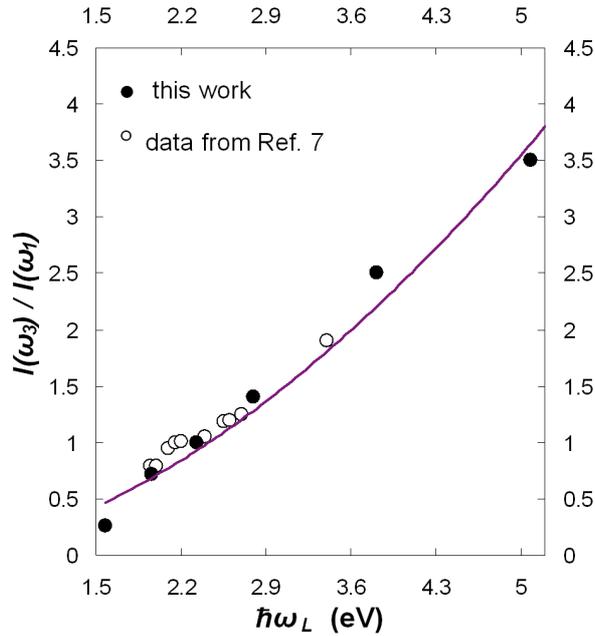

Fig. 7 - The intensity ratio of $I(\omega_3)/I(\omega_1)$ vs. the laser excitation energy $\hbar\omega_L$ for *trans*-$(CH)_x$ inclusions in *a*-C:H. Solid line is a theoretical calculation performed using the amplitude mode formalism [7].

*3.2. Identification of sp-hybridized inclusions in a-C:H*

The observation that contributions belonging to *sp*-hybridised species could be deducted from the main XPS $C_{1s}$ core level electron spectra were first reported by Sergushin *et al.* [36] for X-ray studies of "carbyne" that suggested a much lower binding energy (BE) level for an *sp*-allotrope, relative to BE for elemental carbon contributions. Figure 8 shows the broad core-level XPS $C_{1s}$ spectra of *a*-C:H; in order to deduct the information about the relative abundance of carbon hybridised fractions in the examined materials, the spectra were decomposition onto three main constituent peaks corresponding to *sp*, *sp²* and *sp³* hybridised states. An *sp* peak was fitted at an assigned BE of 283.5 eV in confirmation with previous reports [26, 35, 36], while *sp²* and *sp³* constituents were fitted restricted to their respective energy separation gap, ΔBE, eV, defined as the difference between the binding energies of *sp³* and *sp²* constituents: ΔBE = BE$_{sp3}$ − BE$_{sp2}$ and with 0.85≤ ΔBE ≤ 0.9 eV [37, 55] the following binding energy positions were obtained for *sp²* at ~284.4 eV and *sp³* at ~285.3 eV. Due to *ex situ* XPS measurements and the exposure of samples to air two secondary peaks were added into the fitting of the main $C_{1s}$ spectra: a single C–O peak at ~286.8 eV and a carbonyl C=O peak at ~288.5 eV. The *sp*/*sp²*/*sp³* value ratio

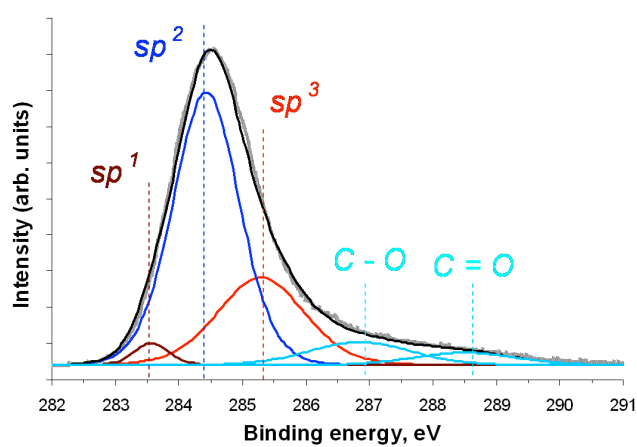

Fig. 8 - XPS core-level $C_{1s}$ spectra of *a*-C:H; contributions for *sp*, *sp²* and *sp³* fractions are shown together with C-O and C=O secondary peaks.

was calculated on the basis of integrating peak areas for the three respective line-functions and was found to be 0.03/0.67/0.30 or, when expressed as a percentage: 3% *sp*, 67% *sp²* and 30% *sp³*; the absolute uncertainty of the measurements was high at

≤1.25% owning to variable ΔBE gap parameter used in the non-linear least squares fitting procedure. The FWHM for *sp*, *sp*$^2$ and *sp*$^3$ were found to be correspondingly ~0.6 eV, ~1.2 eV and ~1.6 eV, to some extent wider that FWHM values reported for general *a*-C:H or *ta*-C:H materials.

The Raman spectra in the range of 1900 – 2200 cm$^{-1}$ are commonly identified with *sp*-hybridised species [24, 25, 38, 39]. Figure 9 shows the RRS of examined *a*-C:H films; the spectra from 532 nm to 325 nm displays only minor perturbations in this range and the *sp* contributions become clearly visible when probed by 244 nm laser. Stability of polyyne and cumulene (*sp*-bonded) species is greatly influenced by the hydrogen environment, and polyynes are known to be significantly more resistant to hydrogen exposure [21]. For that reason we consider polyynes to be prevailing species in the studied *a*-C:H. Polyynes have been theoretically predicted [23, 56] to be more energetically stable than cumulenes, and such preferential stability have been evidenced experimentally [22, 57].

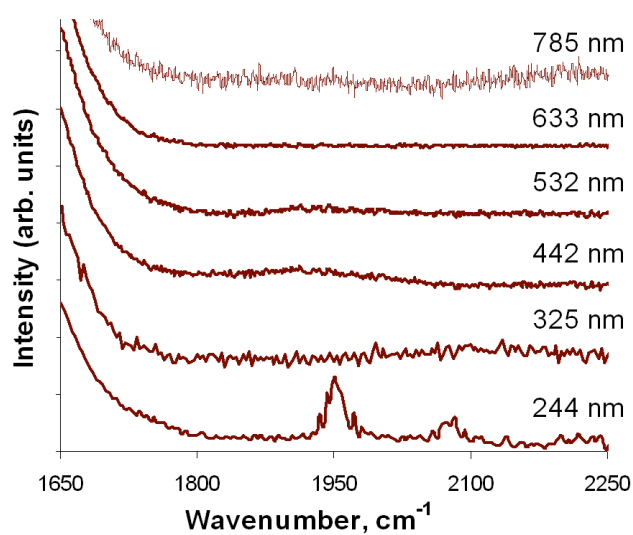

Fig. 9 - Resonant Raman spectra of *sp*-hybridised (polyyne) segments in *a*-C:H.

Recent works by Tabata *et al.* [58, 59] substantiated the assignment of Raman frequency modes centred at around 2000 cm$^{-1}$ to hydrogen capped short H–(CC)$_n$–H polyynic chains with $n = 8 - 18$, as did other workers on this subject [60, 61]. The assignment of the peaks at ~1950 cm$^{-1}$ and 2070 cm$^{-1}$ which appeared in the 244 nm spectrum to fixed-length polyynes is unjustified in our case, since the great majority of published work on the subject considers *sp* inclusions as completely detached. Satisfactory explanation for observation of polyynes exclusively under UV excitation (Fig. 8) could no longer be regarded exclusively resulting from the resonance enhancement of apparently smaller cross section area of one dimensional *sp*-bonded atoms, as we considered previously [38] relying on D'Urso *et al.* [19] findings, which suggested nearly an exponential increase of a combined *sp*–*sp*$^2$ probing signal with respect to increasing $\hbar\omega_L$. Recently released *ab initio* calculations within density functional theory by Ravagnan *et al.* [62] showed that experimental Raman spectra in the range of 1900 – 2200 cm$^{-1}$ for torsionally strained *sp*-nanowires stabilised by *sp*$^2$ and *sp*$^3$ terminations were highly sensitive to strain; changes to the relative orientation of the terminations were found to affect the strain, which subsequently, modulates the electronic states of the nanowires and Raman signal. The appearance of *sp*-bonded species under UV excitation only (VDOS probing) could be attributed to aligned *sp*-hybridised atoms bridging a nanometric gap on *sp*$^2$–*sp*$^3$ matrix; such *sp* chains are most likely to be end-terminated by an *sp*$^2$ or an *sp*$^3$ hybridised fragment; *sp*-stabilisation could be achieved in a form of end-termination or bridging.

Figure 10 illustrates C–H stretching band spectra in the range of 3400 – 2700 cm$^{-1}$ taken from *a*-C:H samples after a baseline correction.

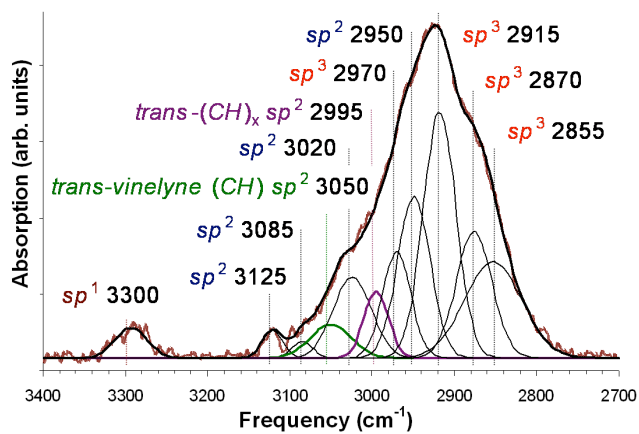

Fig. 10 - Decomposed IR stretching vibrations spectra of an *a*-C:H film. The constituent bonding groups are: $sp^1$(3300) CH, $sp^2$(3125) =C-H unsat/A/asym, $sp^2$(3085) =CH$_2$ unsat/O/asym, $sp^2$(3050) =C-H sat/A/asym, $sp^2$(3020) *trans*-vinelyne (CH) sat/O/sym, $sp^2$(2995) *trans*-(CH)$_x$ sat/O/sym, $sp^3$(2970) -CH$_3$ sat/O/asym, $sp^2$(2950) =CH$_2$ sat/O/asym, $sp^3$(2915) =CH, =CH$_2$ sat/O/asym, $sp^3$(2870) -CH$_3$ sat/O/sym, $sp^3$(2855) =CH$_2$ sat/O/sym.

The configuration of constituent groups is shown in abbreviated form as: saturated (sat); unsaturated (unsat); aromatic (A); olefinic (O); symmetric (sym); and asymmetric (asym). The quantitative [5, 6, 32, 63] decomposition reveals contributions from the *sp*-bonded species at 3000 cm$^{-1}$; the amount of *sp*-hybridised inclusions estimated using Liu *et al.* [32] calculations never exceeded 3% for all films deposited.

The *trans*-(CH)$_x$ mode is visible at 2995 cm$^{-1}$, however the infrared absorption of this mode is expected to be significantly lower than other *C–H* modes fitted, as compared to free molecules. The relative intensity and FWHM of this mode was found closely related to hydrogen content in the films [13]; the mode contribution for the films studied was between 7 to 10%. The stretching mode corresponding to *trans*-vinelyne was found at approximately 3020 cm$^{-1}$ [50]. The amount of *sp*-bonded species in *a*-C:H was found to some extent related to hydrogen content in the films, the ion energy during deposition and deposition temperature; the amount of *trans*-(CH)$_x$ and PPV inclusions was strongly influenced by plasma density and electron temperature.

## 4. Summary and conclusions

In summary, we have performed the RRS investigation on ICP fabricated *a*-C:H films and have demonstrated that the films host *trans*-(CH)$_x$ segments of various conjugation length, poly(*p*-phenylene vinylene) chains as evidenced by the 1175 cm$^{-1}$ Raman mode and, a small fraction of *sp*-hybridised carbon species as evidenced under UV excitations. We provided the theoretical basis for arguing that at various laser excitation energies the changes in Raman spectra features for *trans*-(CH)$_x$ segments included in the *a*-C:H matrix are practically identical to the changes observed in bulk *trans*-polyacetylene. This leads to reliable identification of *trans*-(CH)$_x$ inclusions in Raman spectra of *a*-C:H or any other complex carbonaceous material and, differentiation between the Raman active modes for *trans*-(CH)$_x$ and PPV.

We observed that relative intensities of *trans*-(CH)$_x$ core $A_g$ modes ($\omega_1$ and $\omega_3$) are essentially equal when probed by excitation energy of green light laser. This is although supported by the theoretical distribution of $I(\omega_3)/I(\omega_1)$ ratio relative to Raman excitation energy, overall exemplifying the approach for *trans*-(CH)$_x$ identification in carbonaceous solids.

The length of *trans*-(CH)$_x$ segments in the films examined was found averaging ~25 (± 5) *C=C* bond length units, with longer chains of up to 120 bond length units probed by NIR and shorter chains of ~8 units probed by UV $\hbar\omega_L$; all *trans*-(CH)$_x$ inclusions irrespective of conjugation length displayed high degree of bonding disorder. We assigned the 1175 cm$^{-1}$ peak to PPV *CC–H* bending mode of the ring and postulated the origin of this mode. *Sp*-hybridised species observed exclusively under UV excitation were identified as short polyynic chains bridging a nanometric gap on $sp^2$–$sp^3$ matrix; and these chains were believed to be end-terminated by an $sp^2$ or an $sp^3$ hybridised fragments.

The presence of *sp*-bonded inclusions in *a*-C:H was confirmed by means of XPS $C_{1s}$ core-electron spectra analysis, while the existence of *sp*, *trans*-(CH)$_x$ inclusions and *trans*-vinelyne segments was verified by FT-IR analysis.

Finally, the inclusions of basic polymeric chains such as long *trans*-(CH)$_x$ and PPV were possible owning to highly ordered $sp^2$ host component of *a*-C:H films. Such $sp^2$ ordering was achieved via fabrication of *a*-C:H films in ICP reactor with high plasma density and low electron temperature compared to conventional DLC deposition systems. A parallel could be drawn with the report by Chen *et al.* [64] where unusual inclusions of silicon based spherical nanocrystallites into DLC matrix was possible due to similar ICP fabrication conditions, whereas the stability of *sp*- hybridised species in *a*-C:H could be attributed to relatively low temperature deposition process.

**Acknowledgements**

The authors are thankful to G. Hope of Griffith University and S. Prawer of the University of Melbourne for assisting with the UV Raman measurements. M.R. is grateful to L. Ravagnan and P. Milani of the University of Milan and members of the Laboratorio Getti Molecolari for valuable discussions. M.R. acknowledges the BAM for the visiting postdoctoral fellowship funding. This work has been supported by the Australian Research Council and Australian Nano TAP program.

**References**

[1]     Bernasconi M, Parrinello M, Chiarotti GL, Focher P, Tosatti E. Anisotropic a-C:H from Compression of Polyacetylene. Physical Review Letters. 1996;76(12):2081.

[2]     Arbuckle GA, MacDiarmid AG, Lefrant S, Verdon T, Mulazzi E, Brivio GP, et al. Optical spectroscopic investigation of segmented trans-polyacetylene. Physical Review B. 1991;43(6):4739.

[3]     Robertson J. Diamond-like amorphous carbon. Materials Science and Engineering: R. 2002;37(4-6):129-281.

[4]     López-Ríos T, Sandré É, Leclercq S, Sauvain É. Polyacetylene in Diamond Films Evidenced by Surface Enhanced Raman Scattering. Physical Review Letters. 1996;76(26):4935-8.

[5]     Dischler B, Bubenzer A, Koidl P. Bonding in hydrogenated hard carbon studied by optical spectroscopy. Solid State Communications. 1983 1983/10;48(2):105-8.

[6]     Piazza F, Golanski A, Schulze S, Relihan G. Transpolyacetylene chains in hydrogenated amorphous carbon films free of nanocrystalline diamond. Applied Physics Letters. 2003;82(3):358-60.

[7]     Ehrenfreund E, Vardeny Z, Brafman O, Horovitz B. Amplitude and phase modes in trans-polyacetylene: Resonant Raman scattering and induced infrared activity. Physical Review B. 1987;36(3):1535 LP - 53.

[8]     Brivio GP, Mulazzi E. Theoretical analysis of absorption and resonant Raman scattering spectra of trans-(CH)x. Physical Review B. 1984;30(2):876.

[9]     Ferrari A, Robertson J. Origin of the 1150 cm-1 Raman mode in nanocrystalline diamond. Physical Review B. 2001;63:121405.

[10]     Pfeiffer R, Kuzmany H, Knoll P, Bokova S, Salk N, Gunther B. Evidence for trans-polyacetylene in nano-crystalline diamond films. Diamond and Related Materials, 13th European Conference on Diamond, Diamond-Like Materials, Carbon Nanotubes, Nitrides and Silicon Carbide. 2003;12(3-7):268-71.

[11]     Kuzmany H, Pfeiffer R, Salk N, Gunther B. The mystery of the 1140 cm-1 Raman line in nanocrystalline diamond films. Carbon, European Materials Research Society 2003, Symposium B: Advanced Multifunctional Nanocarbon Materials and Nanosystems. 2004;42(5-6):911-7.

[12]     Michaelson S, Ternyak O, Hoffman A, Lifshitz Y. Hydrogen incorporation processes in nanodiamond films studied by isotopic induced modifications of Raman spectra. Applied Physics Letters. 2006;89(13):131918.

[13]     Teii K, Ikeda T, Fukutomi A, Uchino K. Effect of hydrogen plasma exposure on the amount of trans-polyacetylene in nanocrystalline diamond films. Journal of Vacuum Science and Technology B. 2006;24(1):263-6.

[14]     Rybachuk M, Hu A, Bell JM. Resonant Raman scattering from polyacetylene and poly(p-phenylene vinylene) chains included into hydrogenated amorphous carbon. Applied Physics Letters. 2008;93(5):051904-3.

[15]     Mullazzi E, Brivio GP, Faulques E, Lefrant S. Experimental and theoretical Raman results in trans polyacetylene. Solid State Communications. 1983;46(12):851-5.

[16]     Brivio GP, Mulazzi E. Absorption and resonant Raman scattering from trans-(CH)x. Chemical Physics Letters. 1983;95(6):555-60.

[17]     Ferrari AC, Robertson J. Resonant Raman spectroscopy of disordered, amorphous, and diamondlike carbon. Physical Review B. 2001;64(7):075414.

[18]     Heeger AJ, Kivelson S, Schrieffer JR, Su W-P. Solitons in conducting polymers. Reviews of Modern Physics. 1988;60(3):781-850.

[19]     D'Urso L, Compagnini G, Puglisi O. sp/sp2 bonding ratio in sp rich amorphous carbon thin films. Carbon. 2006;44(10):2093-6.

[20]     Heimann RB, Kleiman J, Salansky NM. A unified structural approach to linear carbon polytypes. Nature. 1983;306(5939):164-7.

[21]     Lenardi C, Barborini E, Briois V, Lucarelli L, Piseri P, Milani P. NEXAFS characterization of nanostructured carbon thin-films exposed to hydrogen. Diamond and Related Materials. 2001 2001/0;10(3-7):1195-200.

[22]     Agostino RG, Caruso T, Chiarello G, Cupolillo A, Pacile D, Filosa R, et al. Thermal annealing and hydrogen exposure effects on cluster-assembled nanostructured carbon films embedded with transition metal nanoparticles. Physical Review B. 2003;68(3):035413-12.

[23]     Springborg M, Kavan L. On the stability of polyyne. Chemical Physics. 1992 1992/12/15;168(2-3):249-58.

[24]     Ravagnan L, Siviero F, Lenardi C, Piseri P, Barborini E, Milani P, et al. Cluster-Beam Deposition and in situ Characterization of Carbyne-Rich Carbon Films. Physical Review Letters. 2002;89(28):285506-4.

[25]     Hu A, Rybachuk M, Lu QB, Duley WW. Direct synthesis of sp -bonded carbon chains on graphite surface by femtosecond laser irradiation. Applied Physics Letters. 2007;91(13):131906.

[26]     Hu A, Griesing S, Rybachuk M, Lu Q-B, Duley WW. Nanobuckling and x-ray photoelectron spectra of carbyne-rich tetrahedral carbon films deposited by femtosecond laser ablation at cryogenic temperatures. Journal of Applied Physics. 2007;102(7):074311-6.

[27]     Heimann RB. Linear finite carbon chains (carbynes): their role during dynamic transformation of graphite to diamond, and their geometric and electronic structure. Diamond and Related Materials. 1994;3(9):1151-7.

[28]     Baughman RH. Dangerously Seeking Linear Carbon. Science. 2006;312(5776):1009-110.

[29]     Varga IK. Multipurpose plasma generator suitable for diamondlike carbon film formation. J Vac Sci Technol A. 1989;7(4):2639-45.

[30]     Rybachuk M, Bell JM. The observation of sp2 fraction disorder using dual wavelength Raman spectroscopy in a-C:H films fabricated using an open inductively coupled plasma reactor. Diamond and Related Materials. 2006;15(4-8):977-81.

[31]     Stoney GG. The Tension of Metallic Films Deposited by Electrolysis. Proceedings of the Royal Society London A. 1909 May 6;82(82):172 - 5.

[32]     Liu S, Gangopadhyay S, Sreenivas G, Ang SS, Naseem HA. Infrared studies of hydrogenated amorphous carbon (a-C:H) and its alloys (a-C:H,N,F). Physical Review B. 1997;55(19):13020.

[33]     Casiraghi C, Ferrari AC, Robertson J. Raman spectroscopy of hydrogenated amorphous carbons. Physical Review B. 2005;72(8):085401.

[34]     Shirley DA. High-resolution X-ray photoemission spectrum of the valence bands of gold. Physical Review B (Solid State). 1972;5(12):4709-14.

[35]     Zhang L, Ma H, Yao N, Lu Z, Zhang B. Growth and field electron emission properties of nanostructured white carbon films. Journal of Vacuum Science & Technology B: Papers from the 19th International Vacuum NanoElectronics conference. 2007;25:545-7.

[36]     Sergushin IP, Kudryavtsev YP, Élizen VM, Sadovskii AP, Sladkov AM, Nefedov VI, et al. X-ray electron and X-ray spectral study of carbyne. Journal of Structural Chemistry. 1977;18(4):553-5.

[37]     Haerle R, Riedo E, Pasquarello A, Baldereschi A. sp[sup 2]/sp[sup 3] hybridization ratio in amorphous carbon from C 1s core-level shifts: X-ray photoelectron spectroscopy and first-principles calculation. Physical Review B. 2002;65(4):045101.

[38]     Hu A, Lu Q-B, Duley WW, Rybachuk M. Spectroscopic characterization of carbon chains in nanostructured tetrahedral carbon films synthesized by femtosecond pulsed laser deposition. The Journal of Chemical Physics. 2007;126(15):154705.

[39]     Hu A, Rybachuk M, Lu Q-B, Duley WW. Femtosecond pulsed laser deposition and optical properties of diamond-like amorphous carbon films embedded with sp-bonded carbon chains. Diamond and Related Materials. 2008;17(7-10):1643-6.

[40]     Prawer S, Nugent KW, Lifshitz Y, Lempert GD, Grossman E, Kulik J, et al. Systematic variation of the Raman spectra of DLC films as a function of sp2:sp3 composition. Diamond and Related Materials. 1996;5(3-5):433-8.

[41]     Benner DC, Rinsland CP, Devi VM, Smith MAH, Atkins D. A multispectrum nonlinear least squares fitting technique. Journal of Quantitative Spectroscopy and Radiative Transfer. 1995;53(6):705-21.

[42]     Brivio GP, Mulazzi E. Theoretical analysis of absorption and resonant Raman scattering spectra of trans-$(CH)_{x}$. Physical Review B. 1984;30(2):876.

[43]     Fitchen DB. Resonance raman results in polyacetylene. Molecular Crystals and Liquid Crystals. 1982;83(1):95 - 108.

[44]   Kuzmany H. Resonance Raman Scattering from Neutral and Doped Polyacetylene. Physica Status Solidi B. 1980;97(2):521-31.

[45]   Lanzani G, Luzzati S, Tubino R, Dellepiane G. Polarized resonant Raman scattering of cis polyacetylene. The Journal of Chemical Physics. 1989;91(2):732-7.

[46]   Mulazzi E. Polarized resonant Raman scattering spectra from stretched trans polyacetylene. Theory. Solid State Communications. 1985;55(9):807-10.

[47]   Baitoul M, Wery J, Buisson J-P, Arbuckle G, Shah H, Lefrant S, et al. In situ resonant Raman and optical investigations of p-doped poly ( p-phenylene vinylene). Polymer. 2000;41:6955–64.

[48]   Tzolov M, Koch VP, Bruetting W, Schwoerer M. Optical characterization of chemically doped thin films of poly p-phenylene vinylene/. Synthetic Metals. 2000;109:85-9.

[49]   Orion I, Buisson J-P, Lefrant S. Spectroscopic studies of polaronic and bipolaronic species in n-doped poly(paraphenylenevinylene). Physical Review B. 1998;57(12):7050.

[50]   Bradley DDC. Precursor-route poy(p-phenylenevinilene): polymer characterisation and control of electronic properties. Journal of Physics D: Applied Physics. 1987;20:1389 - 410.

[51]   Lichtmann LS, Fitchen DB, Temkin H. Resonant Raman spectroscopy of conducting organic polymers. (CH)x, and an oriented analog. Synthetic Metals. 1980;1(2):139-49.

[52]   Kuzmany H. The particle in the box model for resonance Raman scattering in polyacetylene. Pure and Applied Chemistry. 1985;57(2):235 - 46.

[53]     Tiziani R, Brivio GP, Mulazzi E. Resonant Raman scattering spectra of trans-(CD)_{x}: Evidence for a distribution of conjugation lengths. Physical Review B. 1985;31(6):4015.

[54]     Vardeny Z, Ehrenfreund E, Brafman O, Horovitz B. Resonant Raman Scattering from Amplitude Modes in trans-(CH)x and -(CD)x. Physical Review Letters. 1983;51(25):2326.

[55]     Díaz J, Paolicelli G, Ferrer S, Comin F. Separation of the sp3 and sp2 components in the C1s photoemission spectra of amorphous carbon films. Physical Review B. 1996;54(11 -15):8064–9.

[56]     Kertesz M, Koller J, Azman A. Ab initio Hartree--Fock crystal orbital studies. II. Energy bands of an infinite carbon chain. The Journal of Chemical Physics. 1978;68(6):2779-82.

[57]     Casari CS, Bassi AL, Ravagnan L, Siviero F, Lenardi C, Piseri P, et al. Chemical and thermal stability of carbyne-like structures in cluster-assembled carbon films. Physical Review B. 2004;69(7):075422-7.

[58]     Tabata H, Fujii M, Hayashi S. Surface-enhanced Raman scattering from polyyne solutions. Chemical Physics Letters. 2006 2006/3/10;420(1-3):166-70.

[59]     Tabata H, Fujii M, Hayashi S, Doi T, Wakabayashi T. Raman and surface-enhanced Raman scattering of a series of size-separated polyynes. Carbon. 2006 2006/12;44(15):3168-76.

[60]     Casari CS, Bassi AL, Baserga A, Ravagnan L, Piseri P, Lenardi C, et al. Low-frequency modes in the Raman spectrum of sp-sp[sup 2] nanostructured carbon. Physical Review B. 2008;77(19):195444-7.

[61]     Ravagnan L, Bongiorno G, Bandiera D, Salis E, Piseri P, Milani P, et al. Quantitative evaluation of sp/sp2 hybridization ratio in cluster-assembled carbon films

by in situ near edge X-ray absorption fine structure spectroscopy. Carbon. 2006;44(8):1518-24.

[62]     Ravagnan L, Manini N, Cinquanta E, Onida G, Sangalli D, Motta C, et al. Sp carbon nanowires experiencing axial torsion. arXiv:09022573. 2009 15 February 2009.

[63]     Rybachuk M, Bell JM. The effect of sp2 fraction and bonding disorder on micro-mechanical and electronic properties of a-C:H films. Thin Solid Films. 2007;515(20-21):7855-60.

[64]     Chen L-Y, Hong FC-N. Diamond-like carbon nanocomposite films. Applied Physics Letters. 2003;82(20):3526-8.

**Figure captions**

Fig. 1 - Resonant Raman spectra of bulk *trans*-(CH)$_x$ at 78 K taken for different laser excitation wavelengths. a) $\omega_L$ = 457.9 nm; b) $\omega_L$ = 514.5 nm; c) $\omega_L$ = 600 nm; d) $\omega_L$ = 676.4 nm, adapted from Ref. 15.

Fig. 2 - Resonant Raman spectra of *a*-C:H at 293 K showing contributions from *trans*-(CH)$_x$ ($\omega_1$ and $\omega_3$) modes, PPV (1175 cm$^{-1}$ mode), and DLC (*D*, *G*, and *T* modes). An asymmetric peak visible at NIR–visible (green) $\hbar\omega_L$ at 950 cm$^{-1}$ is the second order *Si*, from Ref. 14.

Fig. 3 a) - Peak dispersion, $\Delta\omega$ and b) peak widths, $\Gamma$ for all constituent peaks as a function of the laser excitation energy $\hbar\omega_L$, from Ref. 14.

Fig. 4 - Changes in the fitted $\omega_1$, $\omega_3$ and *1175* cm$^{-1}$ bands (intensity magnified by a factor of 5) relative to the laser excitation energy $\hbar\omega_L$ in the spectra of *a*-C:H. Light dotted line over $\omega_1$ and $\omega_3$ bands denotes the $I(\omega_3)/I(\omega_1)$ trend.

Fig. 5 a) - Evolution of relative intensities of the *1175* cm$^{-1}$ peak, $I(1175)$ and *trans*-(CH)$_x$ contributions, $I\Sigma(\omega_1, \omega_3)$ and b) relative intensities of $\omega_1$, $I(\omega_1)$ and $\omega_3$, $I(\omega_3)$ peaks as a function of the laser excitation energy $\hbar\omega_L$.

Fig. 6 - Calculated absorption spectra from long and short chains constituting *trans*-$(CH)_x$ samples, adopted from Ref. 16.

Fig. 7 - The intensity ratio of $I(\omega_3)/I(\omega_1)$ vs. the laser excitation energy $\hbar\omega_L$ for *trans*-$(CH)_x$ inclusions in *a*-C:H. Solid line is a theoretical calculation performed using the amplitude mode formalism [7].

Fig. 8 - XPS core-level $C_{1s}$ spectra of *a*-C:H; contributions for $sp$, $sp^2$ and $sp^3$ fractions are shown together with C-O and C=O secondary peaks.

Fig. 9 - Resonant Raman spectra of *sp*-hybridised (polyyne) segments in *a*-C:H.

Fig. 10 - Decomposed IR stretching vibrations spectra of an *a*-C:H film. The constituent bonding groups are: $sp^1$(3300) *CH*, $sp^2$(3125) *=C-H* unsat/A/asym, $sp^2$(3085) *=CH₂* unsat/O/asym, $sp^2$(3050) *=C-H* sat/A/asym, $sp^2$(3020) *trans*-vinelyne (CH) sat/O/sym, $sp^2$(2995) *trans*-$(CH)_x$ sat/O/sym, $sp^3$(2970) *-CH₃* sat/O/asym, $sp^2$(2950) *=CH₂* sat/O/asym, $sp^3$(2915) *=CH, =CH₂* sat/O/asym, $sp^3$(2870) *-CH₃* sat/O/sym, $sp^3$(2855) *=CH₂* sat/O/sym.